\documentclass[conference]{IEEEtran}
\IEEEoverridecommandlockouts
\usepackage{tabularx}             

\usepackage{cite}
\usepackage{amsmath,amssymb,amsfonts}
\usepackage{algorithmic}
\usepackage{graphicx}
\usepackage{textcomp}
\usepackage{xcolor}
\usepackage{breqn}
\def\BibTeX{{\rm B\kern-.05em{\sc i\kern-.025em b}\kern-.08em
    T\kern-.1667em\lower.7ex\hbox{E}\kern-.125emX}}
\begin{document}

\title{Optimized Cascaded Position Control of BLDC Motors Considering Torque Ripple\\
%{\footnotesize \textsuperscript{*}Note: Sub-titles are not captured for https://ieeexplore.ieee.org  and
}

\author{\IEEEauthorblockN{Mohammad Vedadi}
\IEEEauthorblockA{
Munich, Germany \\
m.vedadi@tum.de}
}

\maketitle

\begin{abstract}
Brushless DC (BLDC) motors are increasingly used in various industries due to their reliability, low noise, and extended lifespan compared to traditional DC motors. Their high torque-to-weight ratio and impressive starting torque make them ideal for automotive, robotics, and industrial applications. This paper explores the multi-objective tuning of BLDC motor controllers, focusing on position and torque ripple. A state-space model of the BLDC motor and the entire control system, including the power stage and control structure, is developed in the Simulink environment. Two common control mechanisms, trapezoidal and Field Oriented Control (FOC), are implemented and optimized. Both mechanisms utilize a cascaded closed-loop position control, providing fair disturbance rejection but requiring challenging tuning of the controllers. To address these challenges, the non-dominated sorting genetic algorithm II (NSGA-II) is used for optimization. This study demonstrates the effectiveness of optimization techniques in enhancing the performance of control systems.
\end{abstract}

\begin{IEEEkeywords}
BLDC motor, cascaded control, torque ripple.
\end{IEEEkeywords}

\section{Introduction}
Brushless DC (BLDC) motors are increasingly being used in various industries due to their significant benefits. Without brushes, these motors are more reliable, produce less noise, and have a longer lifespan compared to traditional DC motors. Furthermore, BLDC motors provide a high torque-to-weight ratio and impressive starting torque. These features make them ideal for applications in automotive, robotics, and industrial sectors. \cite{b1}.

The cascade control structure has gained popularity in collaborative robotics and automotive systems, including clutch mechanisms \cite{b2}. Compared to single-loop control, cascade control of BLDC motors enhances system performance. This approach is particularly effective in robotics, providing precise and responsive control, especially in multi-dimensional control scenarios \cite{b3}.

Controlling a BLDC motor is challenging because of its non-linear characteristics. Although various control methods have been proposed, PID controllers remain widely used in practical applications. Their popularity in numerous industries comes from their simplicity and low implementation cost. However, achieving optimal performance with PID controllers requires precise parameter tuning [4].

In recent years, various Evolutionary Algorithms (EAs) have been developed to enhance PID controller optimization. Techniques like Simulated Annealing (SA), and Gravitational Search Algorithm (GSA) have been suggested to improve the dynamic performance of BLDC motors [5-6]. These algorithms typically use the BLDC motor's transfer function, a linear approximation of the actual model, for optimization. Although this approach is straightforward, it doesn't always ensure optimal parameter tuning. Better results can be achieved by applying a search algorithm on a DSP with a real setup. However, this method is not feasible for applications with mechanical constraints, such as robotic arms [7].

In cascaded position control, the primary goal is to reach the target position. For BLDC motor control, it is essential to achieve and maintain a smooth torque response and speed with minimal ripple. This ensures excellent performance in both speed and position control drives [8]. In approaches based on EAs and other optimization methods, the main focus is on reducing position error, often neglecting speed response.

The aim of this work is to study the multi-objective tuning of BLDC motor controller based on position and torque ripple. In this paper, the state-space model of the BLDC motor is built in Simulink environment, as well as the whole control system, including the power stage and control structure. Two very common control mechanisms of BLCD motor, trapezoidal, and Field Oriented Control (FOC), are implemented and optimized. The cascaded closed-loop position is used in both control mechanisms to perform a random trajectory of position. The cascaded controller provides fair disturbance rejection, with the cost of more challenging tuning of the PID position controller and PI speed controller. To tackle the optimization problem, the nondominated sorting genetic algorithm II (NSGA-II) is used in this study.

\section{Modeling and Control Approaches}
\subsection{BLDC Motor Modelling}\label{AA}

\begin{figure}[htbp]
\centerline{\includegraphics[height=5.7cm]{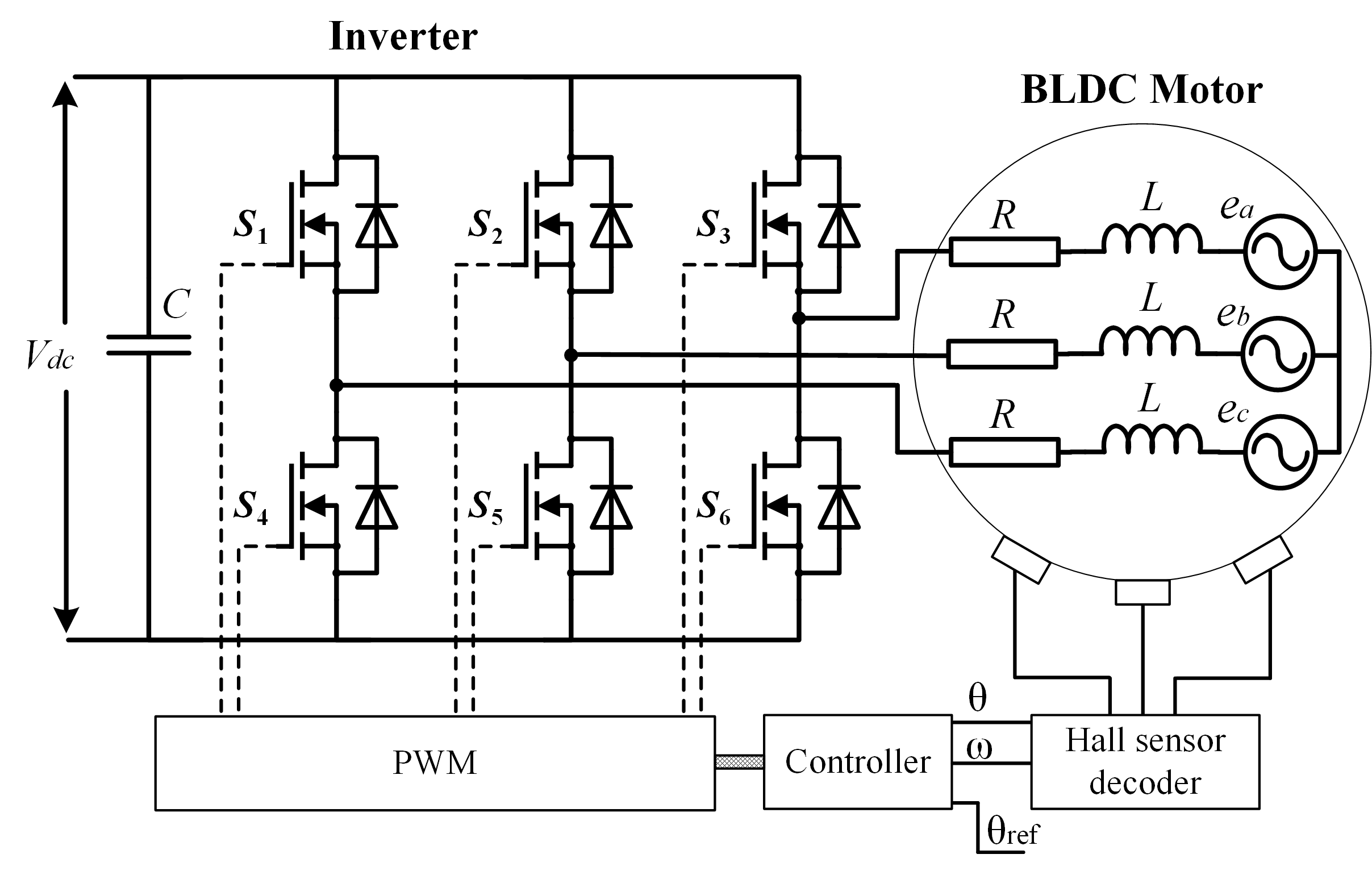}}
\caption{Equivalent circuit of BLDC motor}
\label{fig}
\end{figure}

The equivalent circuit diagram of a BLDC motor is shown in Figure 1. From this diagram, the fundamental equation for the phase voltages ($u_A$, $u_B$, and $u_C$) can be expressed in matrix form as follows:
%\documentclass{article}
%\usepackage{amsmath}  % For math environments

%\begin{document}

\begin{equation}
\begin{aligned}
\begin{bmatrix}
u_A \\    
u_B \\
u_C
\end{bmatrix}
&=
R
\begin{bmatrix}
i_A \\    
i_B \\
i_C
\end{bmatrix} 
+ (L \text{--} M) \frac{d}{dt} 
\begin{bmatrix}
i_A \\    
i_B \\
i_C
\end{bmatrix} \\
&\quad +
\begin{bmatrix}
e_A \\    
e_B \\
e_C
\end{bmatrix}
\end{aligned}
\label{eq:matrix_equation}
\end{equation}

%\end{document}

In this context,R denotes the resistance of the motor windings. The symbols L and M represent the self-inductance and mutual inductance of the stator windings, respectively. The currents in the windings are represented by $i_A$, $i_B$, and $i_C$, while the back electromotive forces (EMFs) of the phases are indicated by $e_A$, $e_B$, and $e_C$.
Ignoring mechanical losses, the entire electromagnetic power is converted into mechanical power. Therefore, the electromagnetic torque can be represented as follows:

\begin{equation}
T_e = \frac{e_a i_a + e_b i_b + e_c i_c}{\omega}
\label{eq:te_equation}
\end{equation}

The angular velocity, \( \omega \), represents the rate of rotational motion. The back electromotive force (EMF) for a phase can be described as \( e = k_e f(\theta) \), where \( k_e \) is the back EMF constant, and \( f(\theta) \) denotes the back EMF waveform as a function of the angular position \( \theta \). Furthermore, the electromagnetic torque can be formulated based on the principle of mechanical torque transfer to the motor shaft.

\begin{equation}
T_e = J \frac{d\omega}{dt} + B\omega + T_L
\label{eq:torque_equation}
\end{equation}

The parameters \( J \) and \( B \) correspond to the rotor's moment of inertia and the frictional coefficient, respectively. The term \( T_L \) signifies the load torque. The relationship between the rotor's electrical speed and its electrical position can be expressed using the following equation.

\begin{equation}
\frac{d\theta}{dt} = \frac{P}{2} \cdot \omega
\label{eq:rotor_speed}
\end{equation}

The symbol \( \theta \) represents the electrical position of the rotor, while \( P \) refers to the number of poles \cite{b9}. The equation for constructing the discrete state-space model can be expressed as follows:

\begin{equation}
x(t+1) = A_d x(t) + B_d u(t)
\label{eq:state_space}
\end{equation}

The system states are represented as \(x = \begin{bmatrix} i_a & i_b & i_c & \omega & \theta \end{bmatrix}^T\), while the input vector is defined as \(u = \begin{bmatrix} V_a & V_b & V_c & T_l \end{bmatrix}^T\). The discrete state matrix is expressed as \(A_d = T_s \cdot A + I\), and the discrete input matrix is given as \(B_d = T_s \cdot B\). Note that the output matrix (\(C\)) is determined as a unity matrix, and the feedforward matrix (\(D\)) is considered as a zero matrix. Thus, the output vector is equal to the state vector (\(Y = x(t)\)). Finally, based on equations (1–4), the standard state matrix and input matrix can be determined [9].

\begin{equation}
A =
\begin{bmatrix}
-\frac{R}{L} & 0 & 0 & -\frac{k_e f_a(\theta)}{L} & 0 \\
0 & -\frac{R}{L} & 0 & -\frac{k_e f_b(\theta)}{L} & 0 \\
0 & 0 & -\frac{R}{L} & -\frac{k_e f_c(\theta)}{L} & 0 \\
\frac{k_e f_a(\theta)}{J} & \frac{k_e f_b(\theta)}{J} & \frac{k_e f_c(\theta)}{J} & -\frac{B}{J} & \frac{P}{2} \\
0 & 0 & 0 & -\frac{P}{2} & 0
\end{bmatrix}
\label{eq:A}
\end{equation}

\begin{equation}
B =
\begin{bmatrix}
\frac{1}{L} & 0 & 0 & 0 & 0 \\
0 & \frac{1}{L} & 0 & 0 & 0 \\
0 & 0 & \frac{1}{L} & 0 & 0 \\
0 & 0 & 0 & -\frac{1}{J} & 0 \\
0 & 0 & 0 & 0 & 0
\end{bmatrix}
\label{eq:B}
\end{equation}

Figure 2 shows the discrete state-space implementation of the BLDC motor in MATLAB/Simulink, as defined by equations (5-7).

\begin{figure}[htbp]
    \centering
    \includegraphics[width=\columnwidth, trim=2.8cm 5.8cm 3.1cm 5.7cm, clip]{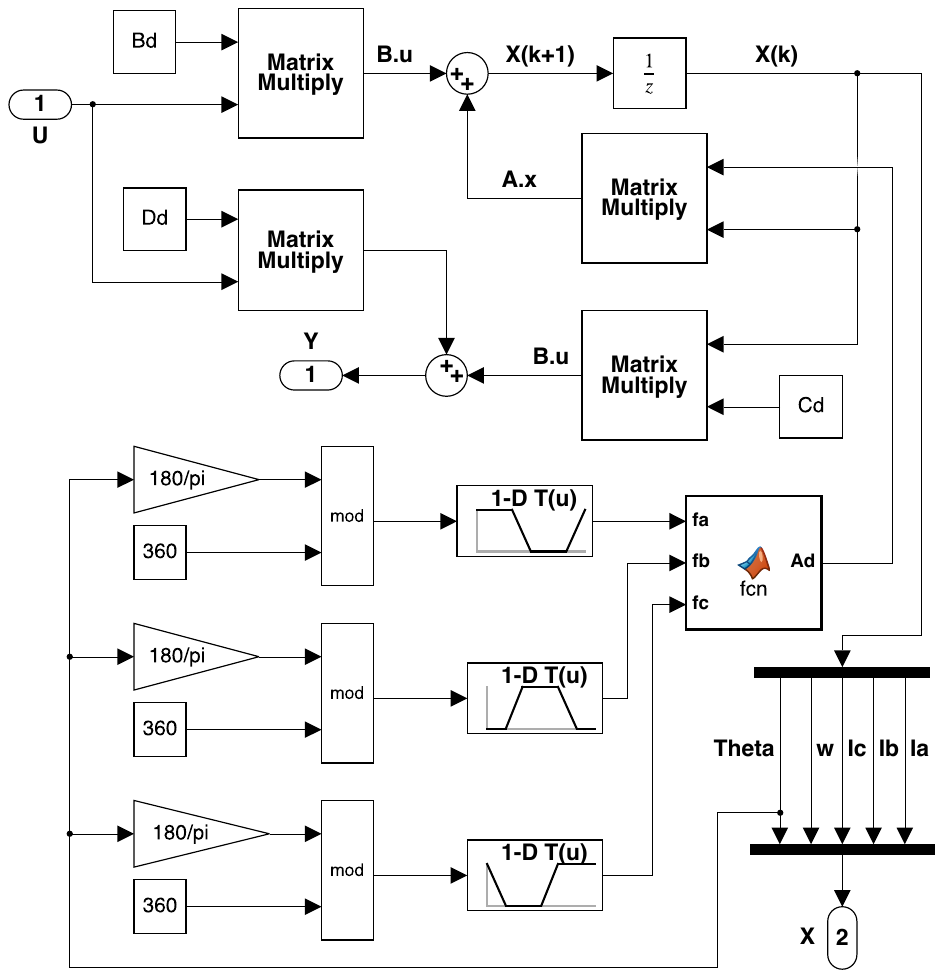}
    \caption{State-space implementation of BLDC motor}
    \label{fig:bldc_motor}
\end{figure}

\subsection{Control Scheme}\label{BB}
This section describes the two control techniques used in our research. The first method, known as six-step commutation or trapezoidal commutation, is a simple way to get the BLDC motor running. In this approach, two of the motor’s three windings are powered at any given time, while the third winding remains inactive. As the motor turns, the current through the windings is switched every 60° of electrical rotation using the VSI. This process results in six distinct positions for the rotor during a full electrical cycle. While Hall sensors are typically used for position feedback in most systems.

Figure 3 presents the Simulink model for trapezoidal commutation. To control the motor, we use a cascaded position control system, which consists of a PI controller for speed in the inner loop and a PID controller for position in the outer loop. This setup modifies the PWM duty cycle, which is sent to the VSI gates.

\begin{figure}[htbp]
    \centering
    \includegraphics[width=\columnwidth, trim=0cm 1.6cm 0cm 1.5cm, clip]{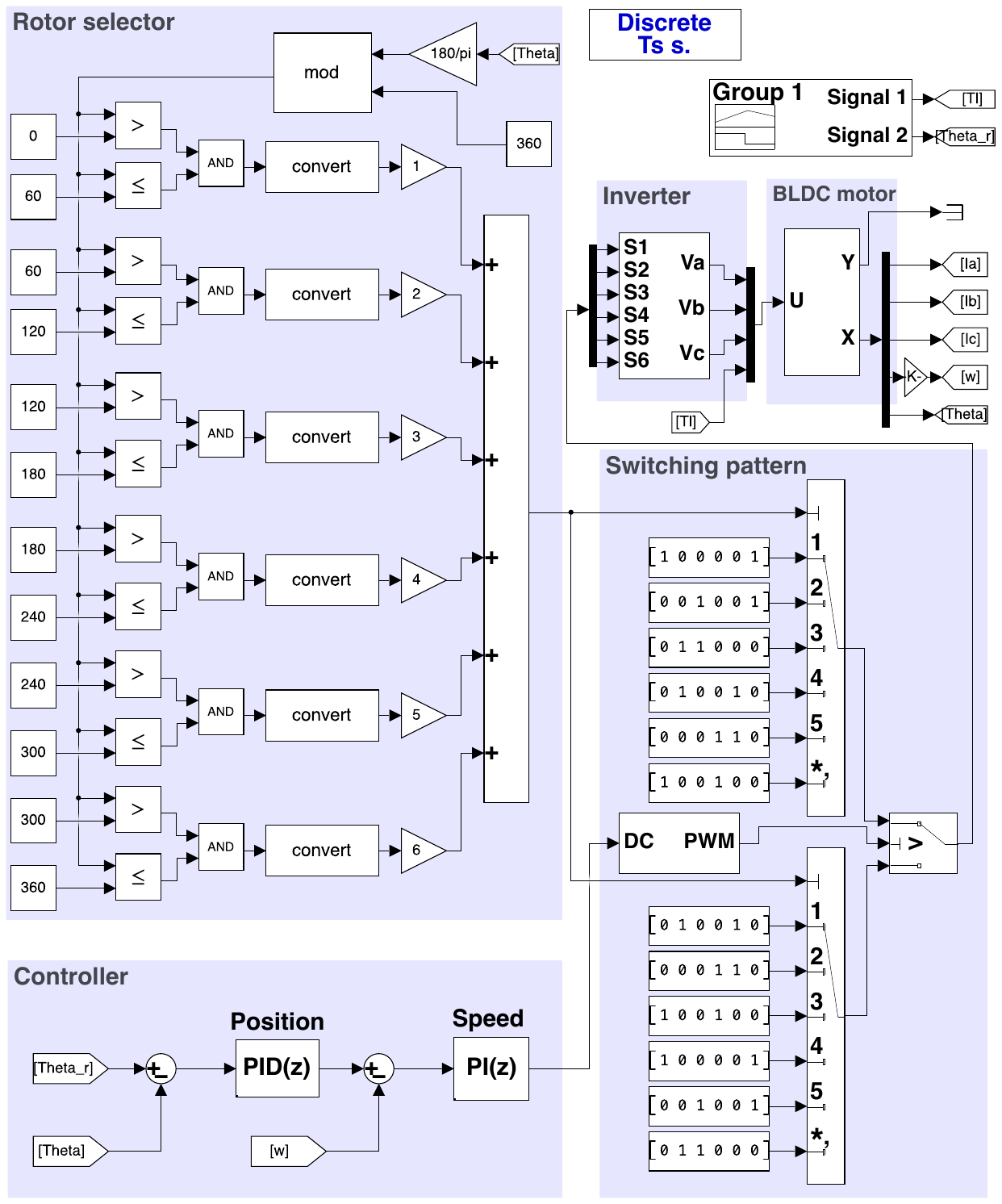}
    \caption{Simulink block diagram of trapezoidal control}
    \label{fig:trapezoidal}
\end{figure}

The Field-Oriented Control (FOC) method in the operation of BLDC motors provides several advantages compared to traditional trapezoidal control. The core principle of FOC involves maintaining a 90° phase difference between the rotor and stator field vectors to achieve optimal torque generation. Since the rotor of a BLDC motor uses permanent magnets, adjustments must be made to the stator's magnetic field. This is accomplished by converting the stator current from a stationary reference frame (abc-frame) to a rotating reference frame (dq-frame), splitting it into components responsible for producing torque and flux.  When operating the motor at or below its rated speed, direct control of the stator flux (the direct axis component) is unnecessary and is set to zero to manage the back EMF effectively. Instead, the quadrature axis component, which dictates torque, is managed through a cascaded control system, as shown in figure 4. FOC provides independent management of current components, as well as reduced torque ripple and lower harmonic distortion in the currents.

\begin{figure}[htbp]
    \centering
    \includegraphics[width=\columnwidth, trim=2.8cm 4cm 2.9cm 3.5cm, clip]{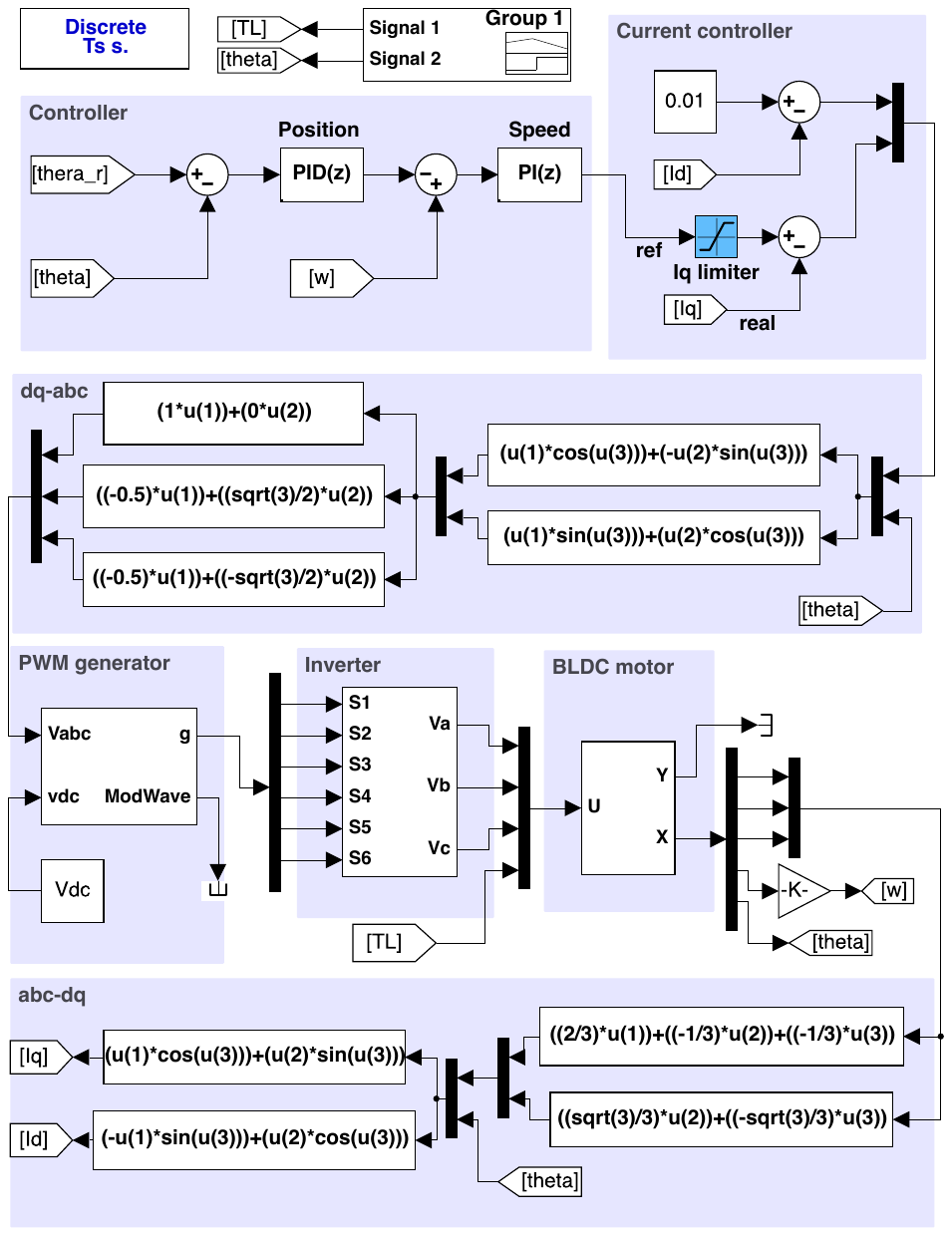}
    \caption{Simulink block diagram of field oriented control}
    \label{fig:FOC}
\end{figure}

\section{Tuning Algorithm}
The optimization problem is addressed using the second version of the Non-Dominated Sorting Genetic Algorithm (NSGA-II) [10]. NSGA-II is an advanced multi-objective evolutionary algorithm that operates on the principle of non-dominated sorting. Among various multi-objective optimization techniques, NSGA-II is recognized for offering an optimal Pareto front, maintaining solution diversity, and achieving rapid convergence.
NSGA-II is utilized to optimize the PID gains for position control in the outer loop of a cascaded control structure. The Integrated Absolute Error (IAE) of position is selected as the first fitness function, as it provides a sensitive measure of system performance, making it well-suited for fitness evaluation. The Total Harmonic Distortion (THD) of torque is chosen as the second fitness function, as it serves as an effective indicator of torque ripple.

A simplified overview of the NSGA-II process for tuning of a controller is summarized as follows:

\begin{enumerate}
    \item Based on the number of initial solutions, a set of control parameters \([k_p, k_i, k_d]\) is generated, referred to as \(P_0\). Where \(k_p, k_i, k_d\) are proportional, integral, and derivative gain of controller, respectively. 
    \item Using the genetic algorithm (GA) operators of crossover and mutation, two additional sets, \(Q_0\) and \(R_0\), are created from \(P_0\).
    \item Calculate the objective function \(f_1\) for each \(k\) using the IAE of position, and calculate \(f_2\) using the THD of torque for all \(k\).
    \item Perform non-dominated sorting of all solutions in \(P_0\), \(Q_0\), and \(R_0\).
    \item Compute the crowding distance for all gain parameters.
    \item Sort the solutions first by their rank (non-dominated front) and then by their crowding distance. Truncate the sorted solutions to match the size of \(P_0\).
    \item If the maximum number of iterations is reached, consider the first Pareto front (\(F_1\)) as the optimal set of solutions. Otherwise, return to Step 2.
\end{enumerate}

Using non-dominated sorting (step 4), the population is divided into fronts iteratively. Solutions are compared based on two objectives, with non-dominated solutions—those not dominated by any other—assigned to the first front (\(F_1\)). Excluding (\(F_1\)), the process is repeated to assign remaining solutions to subsequent fronts (\(F_2\),\(F_3\), etc.) until all solutions are sorted.

Crowding distance is a metric used to prioritize solutions that are more dispersed within the search space. This ensures that solutions with greater diversity are favored during the optimization process. The crowding distance is calculated by measuring the relative distance between adjacent solutions and comparing it to the overall range of the solution set. The following equation illustrates the calculation of the crowding distance:

\begin{equation}
CD = \frac{\left|f_1^{(i+1)} - f_1^{(i-1)}\right|}{\left|f_1^{\text{max}} - f_1^{\text{min}}\right|} 
+ \frac{\left|f_2^{(i+1)} - f_2^{(i-1)}\right|}{\left|f_2^{\text{max}} - f_2^{\text{min}}\right|}
\end{equation}

Figure 5 illustrates the overall procedure of NSGA-II, including the steps for population initialization, non-dominated sorting, and truncation.

\begin{figure}[htbp]
    \centering
    \includegraphics[width=\columnwidth, trim=1.9cm 22.9cm 8.3cm 2.4cm, clip]{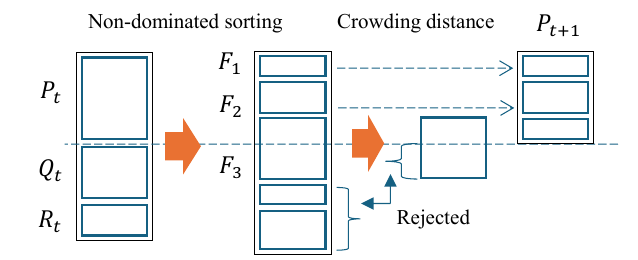}
    \caption{The procedure of NSGA-II}
    \label{fig:nsga}
\end{figure}

\section{Simulation ans result}
%\subsection{System description}
In our study, we utilize a brushless gimbal motor (GBM2804H-100T). The parameters of the BLDC motor are summarized in Table I. For the modeling of the BLDC motor, we have considered an operating voltage of 18V and a switching frequency of 20 kHz.
The optimization of the cascaded position controller is carried out in two distinct stages. Initially, the outer loop's PID position controller is excluded, focusing solely on optimizing the PI speed controller. This is achieved using the methodology outlined in [10]. Once the optimal gains for the PI controller are determined, the second stage begins, wherein the PID position controller is optimized using the NSGA-II algorithm.

\begin{table}[htbp]
\caption{Motor Parameters}
\begin{center}
\renewcommand{\arraystretch}{1.2} % Adjust row height for better spacing
\setlength{\tabcolsep}{4pt}       % Adjust column spacing
\begin{tabularx}{\linewidth}{|X|c|X|c|}
\hline
\textbf{Parameter} & \textbf{Value} & \textbf{Parameter} & \textbf{Value} \\
\hline
Stator resistance ($R$) & 5.6 ($\Omega$) & Moment of inertia ($J$) & 480 ($nN.m.s^2$) \\
\hline
Stator inductance ($L$) & 0.92 ($mH$) & Number of Poles ($P$) & 7 \\
\hline
Back EMF coefficient ($k_e$) & 0.047 ($V.s/rad$) & Rated voltage ($V_{dc}$) & 12 ($V$) \\
\hline
Friction coefficient ($B$) & 550 ($nN.m.s$) & Torque coefficient ($k_t$) & 0.07 ($N.m/A$) \\
\hline
\end{tabularx}
\label{tab:motor_params}
\end{center}
\end{table}

\vspace{0.25\baselineskip} % Adds half a line space

NSGA-II has a computational complexity of $O(\textnormal{M} N^2)$, where $M$ is the number of objectives and $N$ is the population size. Here, $M$ is 2, and a population size of 10 is used to keep the complexity low. Online tuning using NSGA-II can be computationally expensive, therefore the proposed method is well suited for offline tuning where system dynamics are stable. A hybrid approach can also be used, by combining offline tuning for baseline gains with online adaptation for fine-tuning.

Figure 6 shows the Pareto Front of a BLDC Motor with Trapezoidal Control, with each solution indexed.

\begin{figure}[htbp]
    \centering
    \includegraphics[width=\columnwidth, trim=0.8cm 7.9cm 0.7cm 8cm, clip]{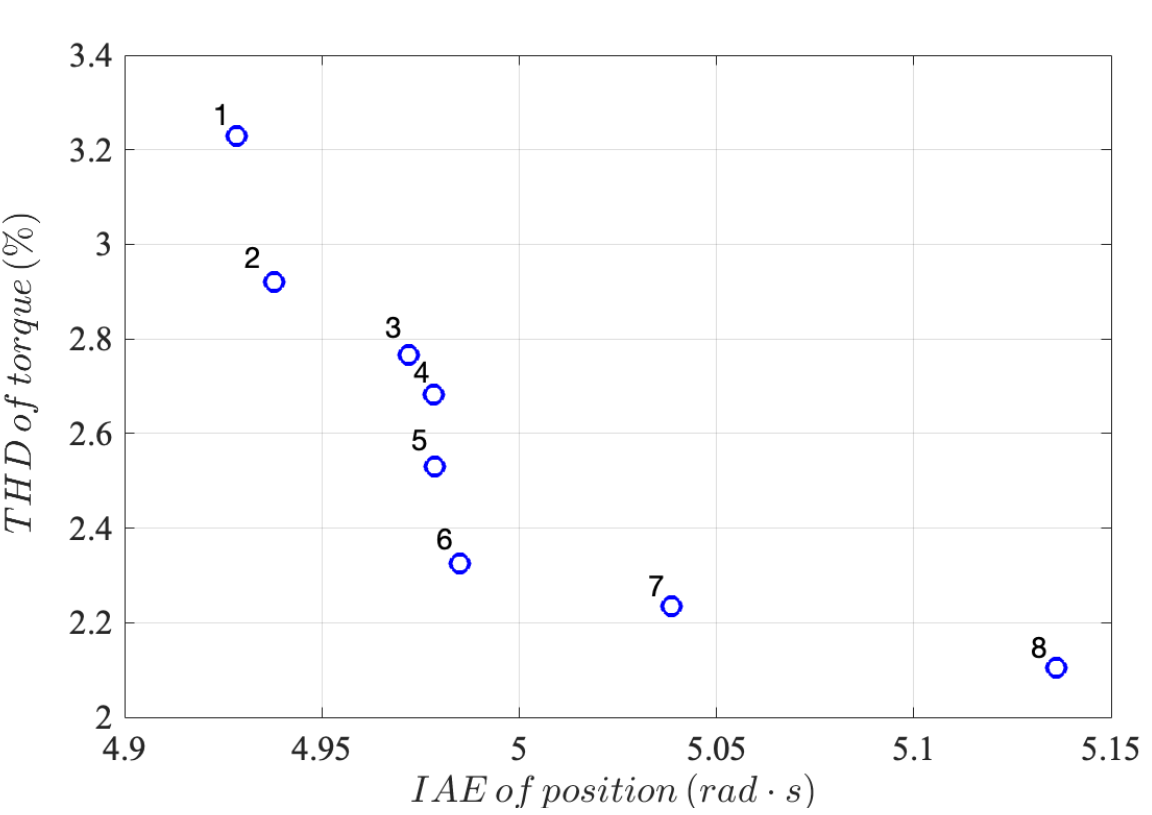}
    \caption{Pareto Front in BLDC Motor with Trapezoidal Control}
    \label{fig:nsga}
\end{figure}

Figure 7 illustrates the Pareto Front for a BLDC Motor utilizing FOC, with each solution distinctly indexed.

\begin{figure}[htbp]
    \centering
    \includegraphics[width=\columnwidth, trim=0.8cm 7.8cm 0.9cm 8.4cm, clip]{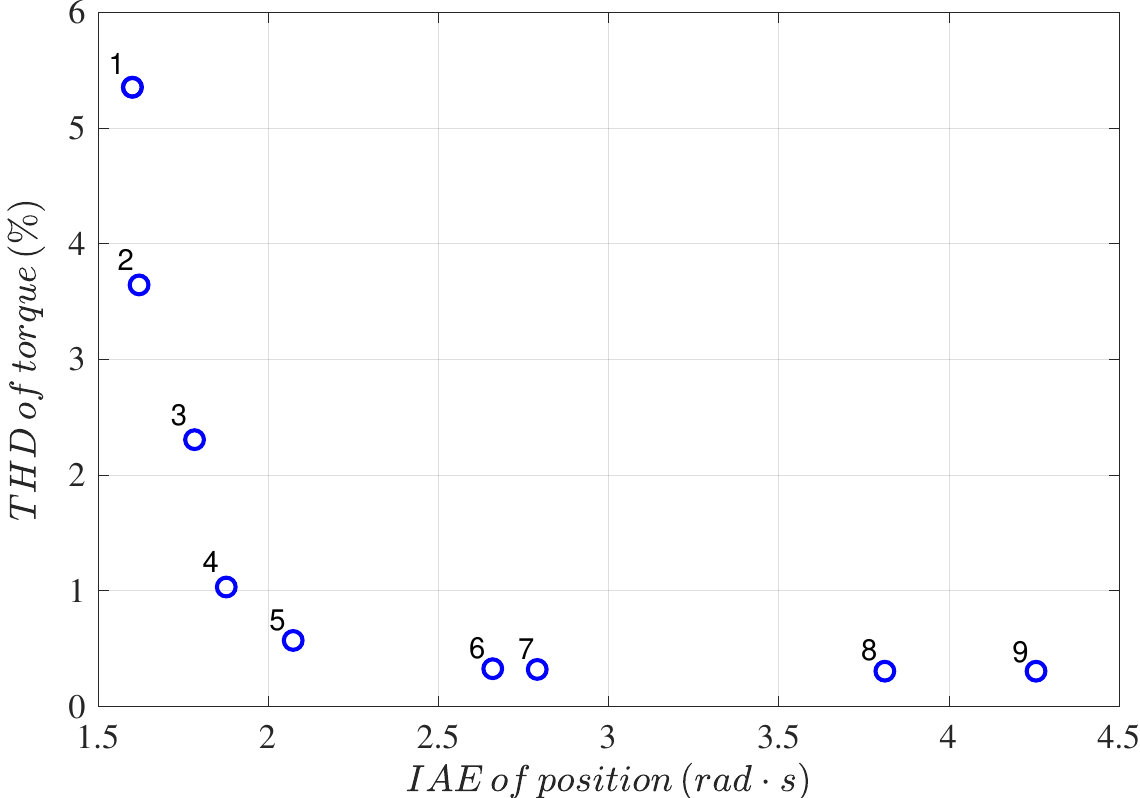}
    \caption{Pareto Front in BLDC Motor with Field-Oriented Control}
    \label{fig:nsga}
\end{figure}
\vspace{0.4\baselineskip}

As expected, FOC demonstrates lower torque distortion compared to the trapezoidal control structure. FOC also achieves superior position control, aligning with anticipated results. It can be observed that FOC provides a broader range of solutions, offering greater flexibility to optimize either position accuracy or torque quality. In contrast, trapezoidal control appears more limited, with fewer options available to minimize torque ripple. The shape of the Pareto front in FOC suggests that optimal solutions balancing both objectives are more readily attainable.

Three solutions were selected from Figure 6 to analyze their performance and gain better insights into the trade-offs between position accuracy and torque ripple in trapezoidal control of a BLDC motor. Boundary solutions and a knee point were selected due to the narrow spread of both fitness values. Figure 8 shows the response position of the solutions.

\begin{figure}[htbp]
    \centering
    \includegraphics[width=\columnwidth, trim=0.7cm 7.65cm 0.6cm 7.7cm, clip]{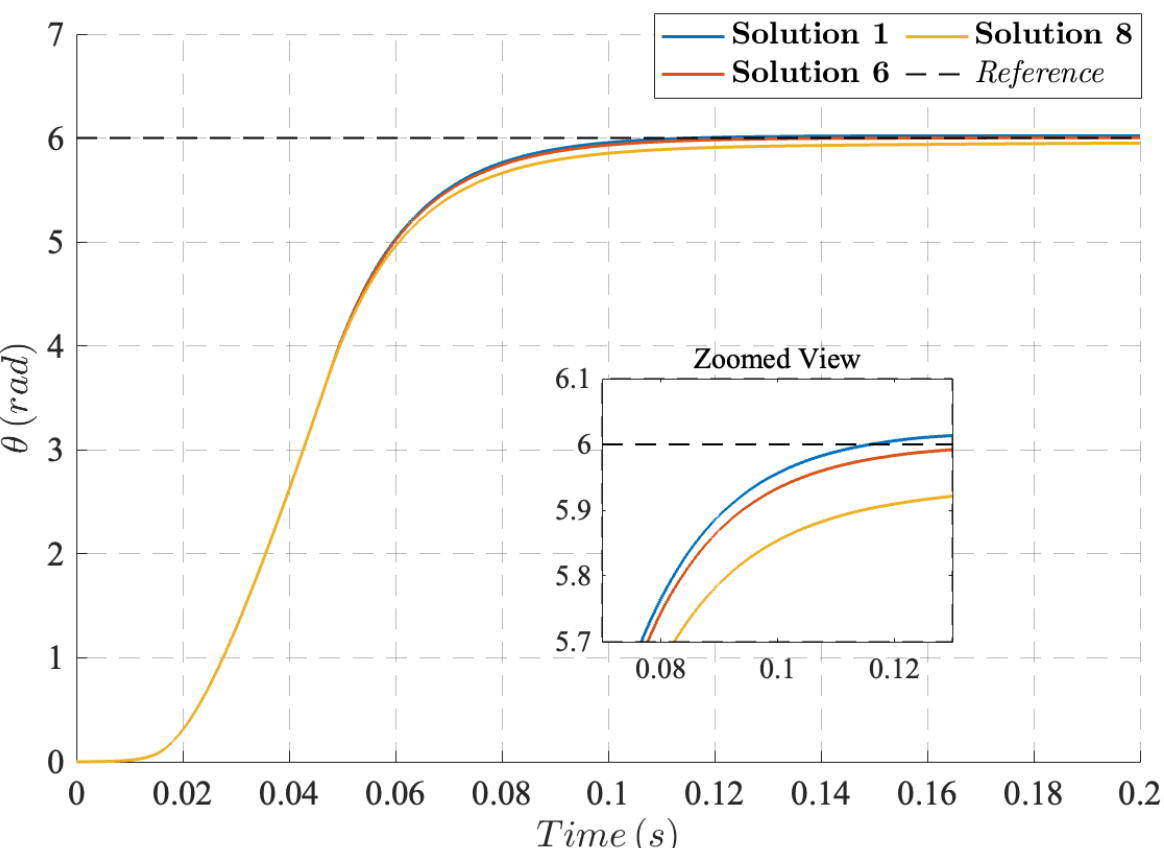}
    \caption{Tracked position response for three solutions in trapezoidal control}
    \label{fig:nsga}
\end{figure}

Solutions 1 and 6 demonstrate optimal performance in terms of rise time and steady-state error. Solution 1, with its faster response, is well-suited for robotics and automation applications where quick positioning is essential. In contrast, Solution 6, which exhibits the lowest steady-state error, is ideal for precision instruments that require high accuracy and minimal deviation from the reference value.

Figure 9 presents the torque profiles for three solutions, highlighting the similarities between Solutions 1 and 6, with Solution 6 exhibiting slightly improved damping after the transition. In precision applications, torque ripple is a critical factor as even minor variations can lead to significant inaccuracies. Conversely, robotics and automation systems can tolerate slightly higher levels of torque ripple without compromising performance. This distinction underscores the importance of selecting the appropriate solution based on the specific requirements of the application.

\begin{figure}[htbp]
    \centering
    \includegraphics[width=\columnwidth, trim=0.7cm 7.6cm 0.8cm 8.1cm, clip]{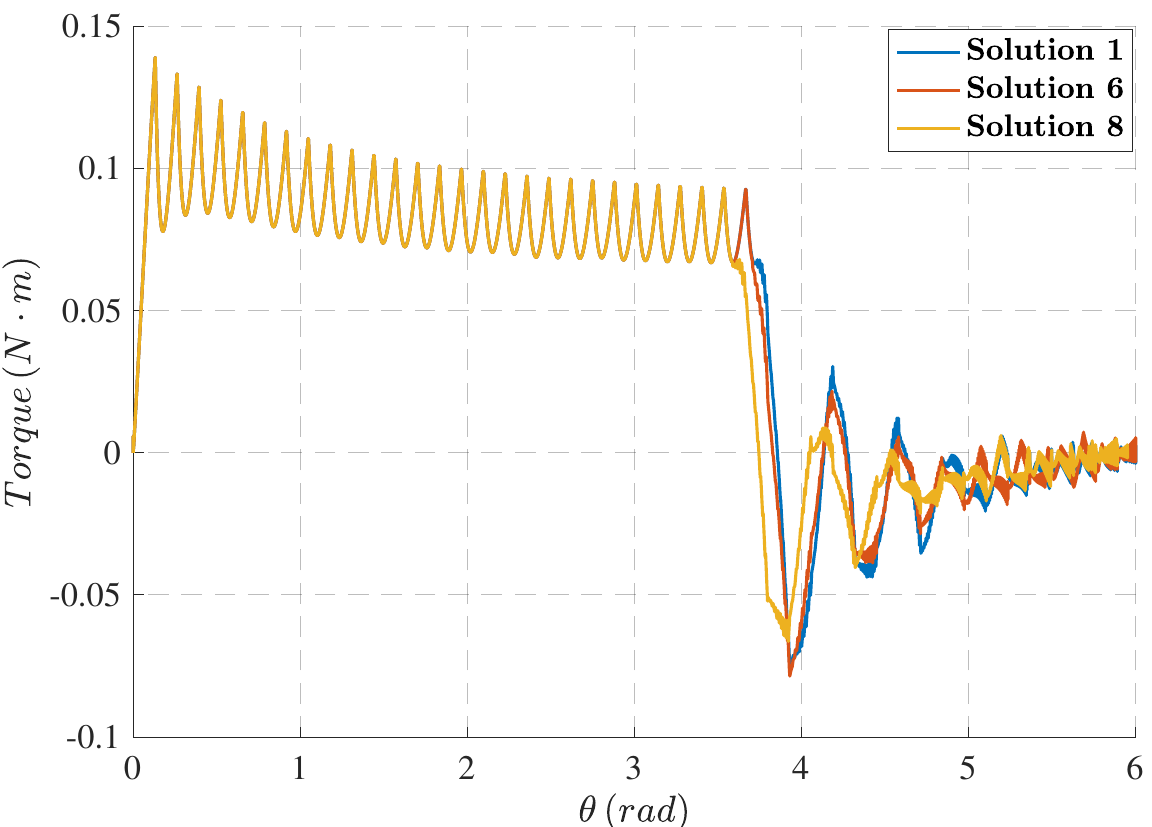}
    \caption{Torque profiles for three solutions in trapezoidal control}
    \label{fig:nsga}
\end{figure}

Similarly, three solutions were identified in Figure 7 to assess performance and examine the trade-offs between position accuracy and torque ripple in the FOC of a BLDC motor. Given the wide spread of fitness values and the clustering of certain solutions with similar performance in one fitness function, boundary solutions were not chosen. Instead, solutions 2, 4, and 6 were selected to represent a diverse range of performance characteristics on the Pareto front.

Figures 10 and 11 illustrate the performance of three selected solutions in FOC. Figure 10 presents the tracked position response over time, highlighting the convergence behavior of each solution with respect to the reference value. Meanwhile, Figure 11 displays the corresponding torque profiles, providing insight into the trade-offs between torque ripple and position accuracy among the different solutions.

The tracked position response follows a trend similar to that observed in trapezoidal control, but with more distinct and noticeable differences between the solutions. While all solutions converge to the reference position with zero steady-state error, they exhibit varying rise times, highlighting differences in dynamic performance.

\begin{figure}[htbp]
    \centering
    \includegraphics[width=\columnwidth, trim=0.3cm 7.6cm 0.8cm 7.7cm, clip]{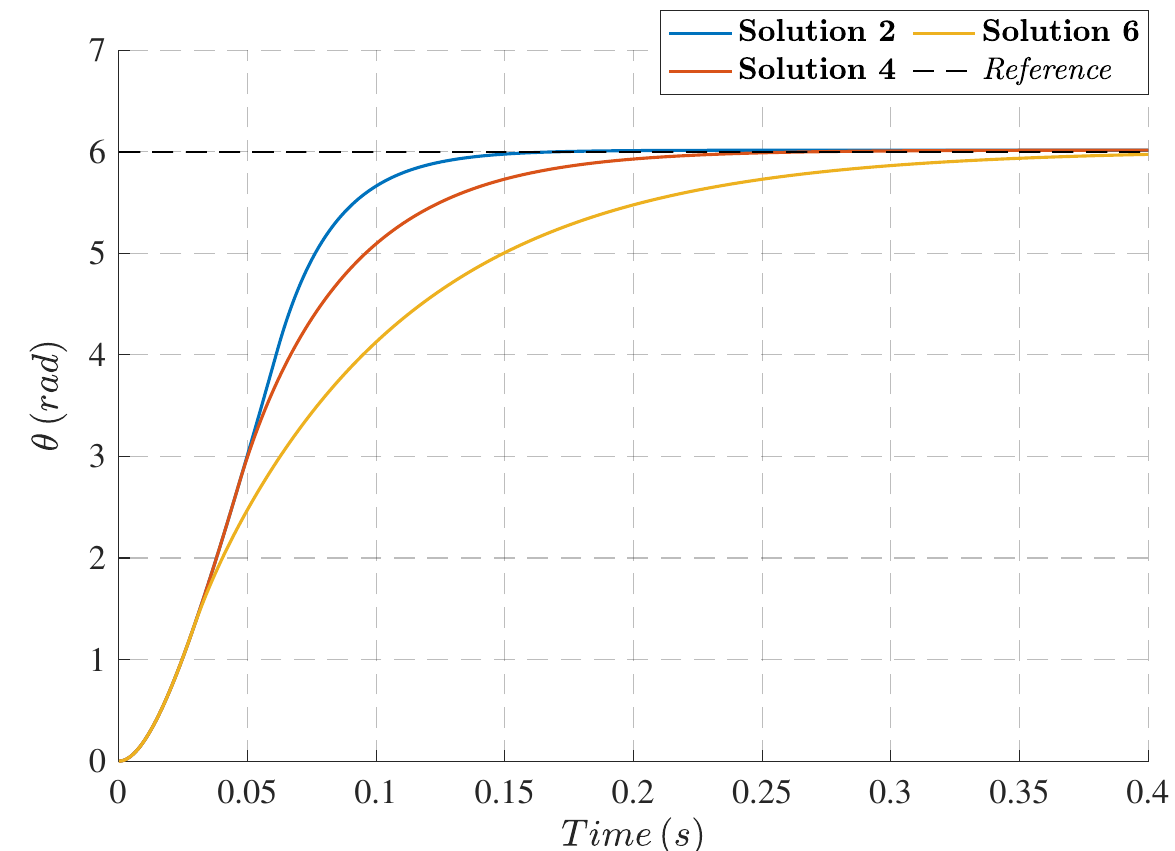}
    \caption{Tracked position response for three solutions in Field_Oriented Control}
    \label{fig:nsga}
\end{figure}

\begin{figure}[htbp]
    \centering
    \includegraphics[width=\columnwidth, trim=0.7cm 7.6cm 0.8cm 7.5cm, clip]{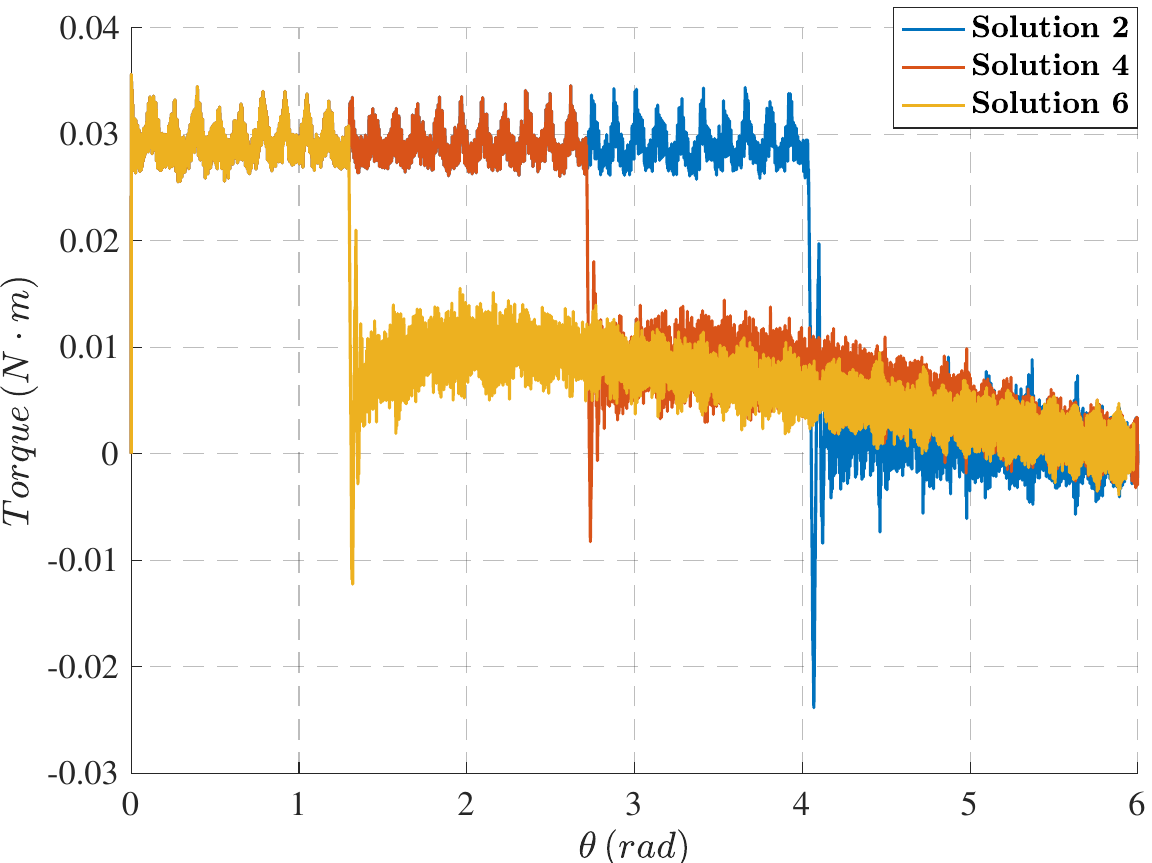}
    \caption{Torque profiles for three solutions in Field-Oriented Control}
    \label{fig:nsga}
\end{figure}

Based on Figure 11, the solutions exhibit higher and more irregular torque ripple compared to the smoother torque profile observed in trapezoidal control (Figure 9). The oscillations in FOC are more noticeable and persistent throughout the operating range.

Each solution offers distinct performance for different applications. Solution 2 converges fastest but has the highest torque ripple, making it suitable for rapid positioning where smooth output is less critical. Solution 4 balances speed and ripple, making it versatile for general use. Solution 6, with the slowest response and minimal ripple, is ideal for applications prioritizing smooth, stable torque and precision.

\section{Conclusion}
The study demonstrates the effectiveness of the optimization technique in enhancing the performance of BLDC motor control systems. The implementation and comparison of trapezoidal and FOC methods highlight the advantages and trade-offs of each approach in terms of position accuracy and torque ripple. FOC outperforms trapezoidal control by offering lower torque ripple and higher flexibility in tuning, as evidenced by the broader and more diverse Pareto front. Trapezoidal control remains a viable option for simpler, faster applications.

The simulation results validate the capability of NSGA-II to optimize PID position controller, achieving precise control over motor position while minimizing torque ripple. The distinct performance of different solutions underscores the importance of multi-objective optimization, allowing system designers to select configurations best suited to specific operational requirements. 

While this study does not explicitly address trade-offs related to energy efficiency or computational latency,  minimizing torque ripple can indirectly improve energy performance. given that NSGA-II is computationally intensive, it is important to note that the proposed  method is intended for offline tuning or hybrid application.

\end{document}